\documentclass[11pt,fleqn]{article}
\usepackage{amssymb,latexsym,amsmath,amsfonts,graphicx, ebezier}

    \advance\textheight by \topskip

    \numberwithin{equation}{section}

    \def\Re{{\rm Re \,}}
    \def\Im{{\rm Im \,}}

    \def\bigO{{\cal O}}

    \def\sech{{\rm sech}}

    \def\P2n{{\rm P}_{{\rm II}}^{(n)}}

    \newtheorem{theorem}{Theorem}[section]

    \newtheorem{Definition}[theorem]{Definition}
    
    \newtheorem{Remark}[theorem]{Remark}
    \newenvironment{remark}{\begin{Remark}\rm}{\end{Remark}}
    \newtheorem{Example}[theorem]{Example}
    
    \newtheorem{Assumptions}[theorem]{Assumptions}

    \newcommand{\e}{\epsilon}

\def\ga{\gamma}
      \def\be{\beta}

    {\rm \trivlist \item[\hskip \labelsep{\bf Proof. }]}%
    {\hspace*{\fill}$\Box$\endtrivlist}
    {\rm \trivlist \item[\hskip \labelsep{\bf Proof}]}%
    {\hspace*{\fill}$\Box$\endtrivlist}

    \DeclareMathOperator*{\Tr}{Tr}

\begin{document}
\title{Critical asymptotic behavior for the Korteweg-de Vries equation and in random matrix theory}
\author{Tom Claeys and Tamara Grava}

\maketitle

\begin{abstract}
We discuss universality in random matrix theory and in the
study of Hamiltonian partial differential equations. We focus
on universality of critical
behavior and we compare results in unitary random matrix ensembles
with their counterparts for the Korteweg-de Vries equation,
emphasizing the similarities between both subjects.
\end{abstract}

\section{Introduction}

It has been observed and conjectured that
the critical behavior of solutions to Hamiltonian perturbations of hyperbolic and elliptic systems of partial
differential equations
near points of gradient
catastrophe is asymptotically independent of the chosen initial data
and independent of the chosen equation \cite{Dubrovin, DGK}. A
classical example of a Hamiltonian perturbation of a hyperbolic
equation which exhibits such universal behavior, is the Korteweg-de Vries (KdV) equation
\begin{equation}\label{KdV}
u_t+6uu_x+\epsilon^2u_{xxx}=0,\qquad \epsilon>0.
\end{equation}
If one is interested in the behavior of KdV solutions in the small
dispersion limit $\epsilon\to 0$, it is natural to study first the
inviscid Burgers' or Hopf equation $u_t+6uu_x=0$. Given smooth
initial data $u(x,0)=u_0(x)$ decaying at $\pm\infty$, the solution
of this equation is, for $t$ sufficiently small, given by the method
of characteristics: we have $u(x,t)=u(\xi(x,t))$, where $\xi(x,t)$
is given as the solution to the equation
\begin{equation}\label{charact}
x=6tu_0(\xi)+\xi.
\end{equation}
It is easily derived from this implicit form of the solution that
the $x$-derivative of $u(x,t)$ blows up at time
$t_c=\frac{1}{\max_{\xi\in\mathbb R}(-6u_0'(\xi))}$, which is called
the time of gradient catastrophe. After this time, the Hopf solution
$u(x,t)$ ceases to exist in the classical sense.  For $t$ slightly
smaller than the critical time $t_c$  the KdV solution starts to
oscillate as shown numerically in \cite{GK}. For $t> t_c$  the KdV
solution develops a train of rapid  oscillations of wavelenght of
order $\epsilon$.
 In general,
the asymptotics for the KdV solution as $\epsilon\rightarrow 0$ can be described in terms of an
equilibrium problem, discovered by Lax and Levermore \cite{LL, LLV}.
The support of the solution of the equilibrium problem,
which depends on $x$ and $t$, consists of a finite or infinite  union of intervals \cite{G, DKMVZ3}, and the endpoints evolve according to the Whitham
equations \cite{W, FFM}.  For $t<t_c$,  the support of the equilibrium problem consists of one interval and  the KdV solution as $\epsilon\rightarrow 0$ is
approximated by the Hopf solution. For $t>t_c$ the support of the equilibrium problem may consists of several intervals and the KdV solution is approximated  as $\epsilon\rightarrow 0$  by Riemann $\theta$-functions \cite{GP, LL, DVZ, V2}.

The $(x,t)$-plane can thus be divided into different regions labeled
by the number of intervals in the support of the Lax-Levermore
minimization problem. Such regions are independent of $\epsilon$ and
depend only on the initial data. Those regions are separated by a
collection of breaking curves where the number of intervals in the
support changes. We will review recently obtained results concerning
the asymptotic behavior of KdV solutions near curves separating a
one-interval region from a two-interval region. The two interval
region corresponds to the solution of KdV being approximated as
$\epsilon \rightarrow 0$ by the Jacobi elliptic function, the one
interval region corresponds to the solution of KdV being
approximated by  the Hopf solution (\ref{charact}).

\medskip

On the space of $n\times n$ Hermitian matrices, one can define
unitary invariant probability measures of the form
\begin{equation}\label{ensemble}\frac{1}{\tilde Z_n}\exp(-N\Tr V(M))dM,\qquad dM=\prod_{i=1}^n d
M_{ii}\prod_{i<j}d \Re M_{ij}d \Im M_{ij},\end{equation} where
$\tilde Z_n=\tilde{Z}_n(N)$ is a normalization constant which depends on the integer $N$ and $V$ is a real
polynomial of even degree with positive leading coefficient. The
eigenvalues of random matrices in such a unitary ensemble follow a
determinantal point process defined by
\begin{equation}
\label{eigenvalues}\frac{1}{Z_n}\prod_{i<j}(\lambda_i-\lambda_j)^2\prod_{i=1}^ne^{-NV(\lambda_i)}d\lambda_i
,\end{equation} with correlation kernel
\begin{equation} \label{kernel}
    K_n(u,v)
        =\frac{e^{-\frac{N}{2}V(u)}
e^{-\frac{N}{2}V(v)}}{u-v}\frac{\kappa_{n-1}}{\kappa_{n}}
(p_n(u)p_{n-1}(v)-p_n(v)p_{n-1}(u)),
\end{equation}
where $p_k$ is the degree $k$ orthonormal polynomial with respect to
the weight $e^{-NV}$ defined by
\[\int_{\mathbb R}p_j(s)p_k(s)e^{-NV(s)}ds=\delta_{jk},\qquad j,k\in\mathbb R,\]
and $\kappa_k>0$ is the leading coefficient of $p_k$. The average
counting measure of the eigenvalues has a limit as $n=N\to\infty$. We
will denote this limiting mean eigenvalue distribution by $\mu_V$.
For a general polynomial external field $V$ of degree $2m$, the
support of $\mu_V$ consists of a finite union of at most $m$
intervals \cite{DKM}. If $V$ depends on one or more parameters, the
measure $\mu_V$ will in general also vary with those parameters.
Critical phenomena occur when the number of intervals in the support
of $\mu_V$ changes. A decrease in the number of intervals can be
caused essentially by three different events:
\begin{itemize}
\item[(i)] shrinking of an interval, which disappears ultimately,
\item[(ii)] merging of two intervals to a single interval,
\item[(iii)] simultaneous merging of two intervals and shrinking of one of
those intervals.
\end{itemize}
Near such transitions, double scaling limits of the correlation
kernel are different from the usual sine or Airy kernel. At a type
(i) transition, the limiting kernel is built out of Hermite
polynomials \cite{Eynard, C, M, BL}, at a type (ii) transition the
limiting kernel is built out of functions related to the Painlev\'e
II equation \cite{BI, CKV}, and at a type (iii) transition the
limiting kernel is related to the Painlev\'e I hierarchy \cite{BMP,
CV2}. Higher order transitions, such as the simultaneous merging
and/or shrinking of more than two intervals, can also take place but
will not be considered here. Rather than on the limiting kernels, we
will concentrate on the asymptotic behavior of the recurrence
coefficients of the orthogonal polynomials, defined by the
three-term recurrence relation
\begin{equation}\label{recur}
sp_n(s)=\gamma_{n+1}p_{n+1}(s)+\beta_np_n(s)+\gamma_np_{n-1}(s).
\end{equation}
The recurrence coefficients contain information about the orthogonal
polynomials and about the partition function $Z_n$ of the
determinantal point process (\ref{eigenvalues}) \cite{BIZ, BI3,
EMcL}. The large $n, N$ asymptotics for the recurrence coefficients
 show remarkable similarities with the
asymptotic behavior for KdV solution $u(x,t,\epsilon)$ as $\epsilon\rightarrow 0$.

\section{Phase diagram for the KdV equation}

We assume throughout this section that the ($\e$-independent)
initial data $u_0(x)$ for the KdV equation are real analytic in a
neighborhood of the real line, negative, have a single local minimum
$x_M$ for which $u_0(x_M)=-1$, and that they decay sufficiently
rapidly as $x\to \infty$ in a complex neighborhood of the real line.
The neighborhood of the real line where $u_0$ is analytic and where
the decay holds should contain a sector $\{|\arg
x|<\delta\}\cup\{|\arg(-x)|<\delta\}.$ In addition certain generic
conditions have to be valid; we refer to \cite{CG1} for details
about those. A simple example of admissible initial data is given by
$u_0(x)=-\sech^2(x)$.

\subsection{Regular asymptotics for the KdV solution}

Before the time of gradient catastrophe
$t_c=\frac{1}{\max_{\xi\in\mathbb R}(-6u_0'(\xi))}$, the asymptotics for the KdV
solution $u(x,t,\e)$ as $\e\to 0$ are given by
\[u(x,t,\e)=u(x,t)+\bigO(\e^2),\]
where $u(x,t)$ is the solution to the Hopf equation with initial
data $u_0(x)$, i.e.\ the implicit solution $u_0(\xi(x,t))$ defined
by (\ref{charact}).  The leading term of the above asymptotic
expansion was obtained in \cite{LL} while the error term was
obtained only recently for a larger class of equations and initial
data in \cite{MR}. Such an expansion  still holds  true after the
time of gradient catastrophe as long as $x$ is outside the interval
where the KdV solution develops oscillations. In the oscillatory
region,  the oscillations for some time $t>t_c$  can be approximated
as $\e\to 0$,  by the  elliptic function
\begin{multline}
\label{elliptic1}
u(x,t,\e)=\beta_1+\beta_2+\beta_3+2\alpha\\
+2\e^2\frac{\partial^2}{\partial
x^2}\log\vartheta\left(\frac{\sqrt{\beta_1-\beta_3}}{2\e
K(s)}[x-2t(\beta_1+\beta_2+\beta_3)-q];\tau\right)+\bigO(\e).
\end{multline} Here
\begin{equation}
\alpha=-\beta_1+(\beta_1-\beta_3)\frac{E(s)}{K(s)}, \quad
\tau=i\frac{K'(s)}{K(s)},\quad
s^2=\frac{\beta_2-\beta_3}{\beta_1-\beta_3},
\end{equation}
where $K(s)$ and $E(s)$ are the complete elliptic integrals of the
first and second kind, $K'(s)=K(\sqrt{1-s^2})$, and
$\vartheta(z;\tau)$ is the Jacobi elliptic theta function. In the
formula (\ref{elliptic1}) the term $\beta_1+\beta_2+\beta_3+2\alpha$
is the weak limit of the solution  $u(x,t,\epsilon)$ of KdV as
$\epsilon\rightarrow 0$ and it was derived in the seminal paper
\cite{LL}. The asymptotic description of the  oscillations  by
theta-function  was obtained in \cite{V2}. A heuristic derivation of
formula (\ref{elliptic1})  without the phase, was first obtained in \cite{GP}.   The
phase $q$ in the argument of the Jacobi elliptic theta function
(\ref{elliptic1})  was derived in \cite{DVZ}. It  depends on
$\beta_1,\beta_2,\beta_3$ and on the initial  data and it was
observed in \cite{GK} that  $q$ satisfies a linear
over-determined system of Euler-Poisson-Darboux type derived in
\cite{FRT2, Getall}.   The negative numbers $\beta_1>\beta_2>\beta_3$ depend
on $x$ and $t$ and solve the genus one Whitham equations \cite{W}.
The complete solution of the Whitham equation for the class of
initial data considered, was derived in \cite{FRT1}.

At later times, the KdV solution can, depending on the initial data, develop multi-phase oscillations which can be described in terms of higher genus Whitham equations \cite{FFM} and in terms of Riemann $\theta$ functions \cite{V2,LL,DVZ}.

The parameters $\beta_1,\beta_2,\beta_3$ can be interpreted in terms
of the endpoints of the support $[0,\sqrt{\beta_3+1}]\cup
[\sqrt{\beta_2+1},\sqrt{\beta_1+1}]$ of the minimizer of the
Lax-Levermore  energy functional \cite{LL, LLV, DVZ}.

A transition from the elliptic asymptotic region to the Hopf region
can happen in three different ways:
\begin{itemize}
\item[(i)] $\beta_1$ approaches $\beta_2$ (shrinking  of an  interval),
\item[(ii)] $\beta_2$ approaches $\beta_3$ (merging  of two  intervals),
\item[(iii)] $\beta_1$,$\beta_2$, and $\beta_3$ approach each other (simultaneous shrinking and merging of intervals).
\end{itemize}
The transitions (i), (ii) and (iii)  will lead to an asymptotic
description of the KdV solution which is similar to the asymptotic
description for the recurrence coefficients of orthogonal
polynomials when the number of intervals in the support of the
limiting mean eigenvalue density of random matrix ensembles changes.
A transition of type (iii) takes place at the point of gradient
catastrophe. In the $(x,t)$ plane  after the time of gradient
catastrophe, the oscillations asympotitcally develop in a $V$-shape
region that does not depend on $\epsilon$ see Figure \ref{figure:
phase KdV}.   At the left boundary (the leading edge), a transition
of type (ii) takes place, and at the right boundary (the trailing
edge) we have a type (i) transition.
 \begin{figure}[t]
    \begin{center}
    \setlength{\unitlength}{1truemm}
    \begin{picture}(100,55)(-5,0)

    \put(50,0){\thicklines\circle*{.8}}
    \put(45,10){\thicklines\circle*{1.5}}
    \put(100,1){$x$}
    \put(51,61){$t$}
    \put(15,0){\vector(1,0){85}}
    \put(50,-3){\vector(0,1){65}}
     \qbezier(45,10)(44,15)(10,48)
    \qbezier(45,10)(41,19)(43,54)

    \put(10,20){Hopf}
    \put(60,20){Hopf}
    \put(27,36){elliptic}

    \put(29,21){(ii)}
    \put(43,21){(i)}
    \put(42,6){(iii)}
    \end{picture}
    \caption{Sketch of the phase diagram for the equilibrium problem associated to the KdV equation. Outside the cusp-shaped region, the support of the Lax-Levermore minimizer consists of one interval; inside of two intervals.
    At the cusp point, we have a type (iii) transition, at the left breaking curve one of type (ii), and at the right breaking curve, one of type (i).}
    \label{figure: phase
KdV}
\end{center}
\end{figure}
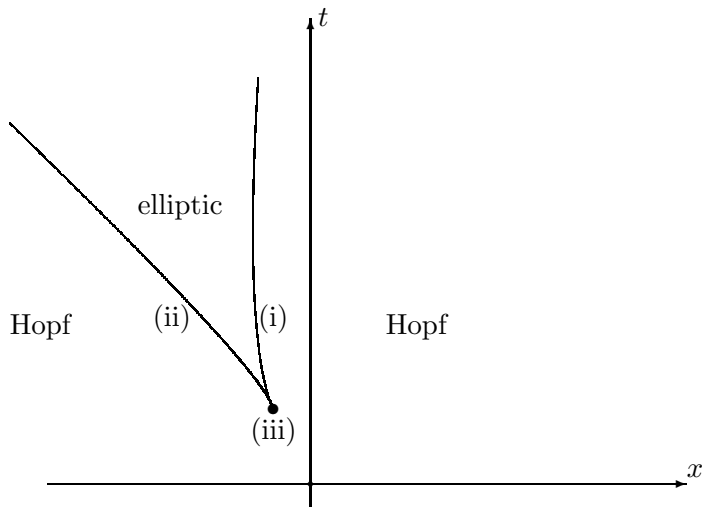
Given $t$ sufficiently short after the time of gradient catastrophe
$t_c$, the leading edge $x^-(t)$ is characterized by the system of
equations
\begin{align}
&\label{leading}x^-(t)=6tu(t)+f_L(u(t)),\\
&\label{leading2}6t+\theta(v(t);u(t))=0,\\
&\label{leading3}\partial_{v}\theta(v(t);u(t))=0,
\end{align}
where $u(t)>v(t)$, $f_L(u)$ is the inverse of the decreasing part of
$u_0(x)$, and $\theta$ is given by
\begin{equation}
\label{theta}
\theta(\lambda;u)=\dfrac{1}{2\sqrt{2}}\int_{-1}^1\dfrac{f'_L(\frac{1+m}{2}\lambda+\frac{1-m}{2}u)d m}{\sqrt{1-m}}.
\end{equation}
This corresponds to the confluent case where  the elliptic solution (\ref{elliptic1}) degenerates formally  to linear oscillations, namely  $\beta_2=\beta_3=v$ and $\beta_1=u$. The trailing edge on
the other hand is characterized by
\begin{align}
&\label{trailing1}x^+(t)=6tu(t)+f_L(u(t)),\\
&\label{trailing2}6t+\theta(v(t);u(t))=0,\\
&\label{trailing3}\int_{u(t)}^{v(t)}(6t+\theta(\lambda;u(t)))\sqrt{\lambda-u(t)}d\lambda=0,
\end{align}
with $u(t)<v(t)$, and $\theta(\lambda;u)$ defined in (\ref{theta}).
In this case we have $\beta_1=\beta_2=v$ and $\beta_3=u$. In this
case the solution (\ref{elliptic1}) degenerates formally to a
soliton.

\subsection{Critical asymptotics for the KdV solution}
\subsubsection{Point of gradient catastrophe}
Near the first break-up time, the KdV solution starts developing
oscillations for small $\e$. These oscillations are modeled by a
Painlev\'e transcendent $U(X,T)$, defined as the unique real smooth
solution to the fourth order ODE
\begin{equation}\label{PI20}
X=T\, U -\left[ \dfrac{1}{6}U^3  +\dfrac{1}{24}( U_{X}^2 + 2 U\,
U_{XX} ) +\frac1{240} U_{XXXX}\right],
\end{equation}
with asymptotic behavior given by
     \begin{equation}
\label{PI2asym}
        U(X,T)=\mp (6|X|)^{1/3}\mp \frac{1}{3}6^{2/3}T|X|^{-1/3}
            +\bigO(|X|^{-1}),
            \qquad\mbox{as $X\to\pm\infty$,}
\end{equation}
for each fixed $T\in\mathbb{R}$. The existence of a pole free solution of (\ref{PI20}) with asymptotic conditions (\ref{PI2asym}) was conjectured in \cite{Dubrovin}) and proved in \cite{CV1}.
 Let us denote $t_c$ for the time of gradient catastrophe, $x_c$ for the point where the $x$-derivative of the Hopf solution blows up, and $u_c=u(x_c,t_c)$.
We take a double scaling limit where we let $\epsilon\to 0$ and
at the same time we let $x\to x_c$ and $t\to t_c$ in such a way
that, for fixed $X,T\in\mathbb R$,
\begin{equation}
\lim\dfrac{x- x_c-6u_c (t-t_c)}{(8k\e^6)^{1/7}}=X, \qquad
\lim\dfrac{6(t-t_c)}{(4k^3\epsilon^4)^{1/7}}= T,
\end{equation}
where
\[
k=-f_L'''(u_c).
\] In this double scaling limit the solution
$u(x,t,\epsilon)$ of the KdV equation (\ref{KdV}) has the following
expansion,
\begin{equation}
\label{expansionu} u(x,t,\e)=u_c
+\left(\dfrac{2\epsilon^2}{k^2}\right)^{1/7} U \left( \dfrac{x-
x_c-6u_c (t-t_c)}{(8k\epsilon^6)^{\frac{1}{7}}},
\dfrac{6(t-t_c)}{(4k^3\epsilon^4)^{\frac{1}{7}}}\right) +O\left(
\epsilon^{4/7}\right).
\end{equation}
 The idea that the solution of KdV near the point of gradient catastrophe can be approximated by the solution of (\ref{PI20}) appeared first in
 \cite{Suleimanov1, Suleimanov2}  and in a more general setting in  \cite{Dubrovin}, and was confirmed rigorously in \cite{CG1}. In \cite{CG4} the correction term of order $\epsilon^{4/7}$ was determined.

\subsubsection{Leading edge}
Near the leading edge, the onset of the oscillations is described by the Hastings-McLeod solution
to the Painlev\'e II equation
 \begin{equation}\label{PII}
 q''(s)=sq+2q^{3}(s).
\end{equation}
The Hastings-McLeod solution is characterized by the asymptotics
\begin{align}
&\label{HM1}q(s)=\sqrt{-s/2}(1+o(1)),&\mbox{ as $s\to -\infty$,}\\
&\label{HM2}q(s)=\mbox{Ai}(s)(1+o(1)), &\mbox{ as $s\to +\infty$,}
\end{align}
where $\mbox{Ai}(s)$ is the Airy function. The leading edge $x^-(t)$
is, for $t$ sufficiently short after $t_c$, determined by the system
of equations (\ref{leading})-(\ref{leading3}). Let us consider a
double scaling limit where we let $\epsilon\to 0$ and at the same
time we let $x\to x^-(t)$ in such a way that
\begin{equation}
\lim \dfrac{x- x^-(t)}{\e^{2/3}}=X \ \in\mathbb R,
\end{equation}
for $t>t_c$ fixed.
In this double scaling limit, the solution $u(x,t,\epsilon)$ of the
KdV equation with initial data $u_0$ has the following
asymptotic expansion,
\begin{equation}
\label{expansionuP}
u(x,t,\e) = u-\dfrac{4\e^{1/3}}{c^{1/3}}q\left[s(x,t,\e)\right]
\cos\left(\frac{\Theta(x,t)}{\epsilon}\right)+O(\e^{\frac{2}{3}}),
\end{equation}
where
\begin{equation}
\label{Theta}
\Theta(x,t)=2\sqrt{u-v}(x-x^-)+2\int_{v}^{u}(f_L'(\xi)+6t)\sqrt{\xi-v}d\xi,
\end{equation}
and
\begin{equation}\label{def c}
c=-\sqrt{u-v}\dfrac{\partial^2}{\partial v^2}\theta(v;u)>0,\qquad
s(x,t,\e)=-\frac{x-x^-}{c^{1/3}\sqrt{u-v}\,\e^{2/3}},
\end{equation}
with $\theta$ defined by (\ref{theta}), and $q$ is the Hastings-McLeod solution to the Painlev\'e II equation.
Here $x^-$ and $v<u$ (each of them depending on $t$) solve the
system (\ref{leading})-(\ref{leading3}).
The above result was proved in \cite{CG2}, confirming numerical
results in \cite{GK}. In \cite{CG2}, an explicit formula for the correction term of order $\epsilon^{\frac{2}{3}}$ was obtained as well.
We remark that a connection between leading edge asymptotics and the Painlev\'e II equation also appeared in \cite{KS}.
\subsubsection{Trailing edge}
The trailing edge $x^+(t)$ of the oscillatory interval (i.e.\ the
right edge of the cusp-shaped region in Figure 2) is determined by
the equations (\ref{trailing1})-(\ref{trailing3}). As $\e\to 0$, we
have, for fixed $y$ and $t$,
\begin{equation}\label{expansion u}
u\left(x^++\frac{\e\ln\e}{2\sqrt{v-u}}y,t,\e\right)=u+2(v-u)\sum_{k=0}^{\infty}\sech^2(X_k)+\bigO(\e\ln^2\e),
\end{equation}
where
\begin{equation}
\begin{split}\label{Xk}
&X_k=\frac{1}{2}(\frac12-y+k)\ln\e-\ln(\sqrt{2\pi} h_k)-(k+\frac12)\ln\gamma,\\
&h_k=\dfrac{2^{\frac{k}{2}} }{\pi^{\frac{1}{4}}\sqrt{k!}},\quad
\gamma=4(v-u)^{\frac{5}{4}}\sqrt{-\partial_v\theta(v;u)},
\end{split}
\end{equation}
and $\theta$ is given by (\ref{theta}) \cite{CG3}.  It should
be noted in this perspective that the KdV equation admits soliton
solutions of the form $a\,\sech^2(bx-ct)$. This means that the last
oscillations of the KdV solution resemble, at the local scale,
solitons.

\section{Phase diagram for unitary random matrix ensembles}
\subsection{Equilibrium problem}
In unitary random matrix ensembles of the form (\ref{ensemble}), the
limiting mean eigenvalue density is characterized as the equilibrium
measure minimizing the logarithmic energy
\begin{equation}
I_V(\mu)=\iint \log\frac{1}{|s-y|}d\mu(s)d\mu(y)+\int V(s)d\mu(s),
\end{equation}
among all probability measures on $\mathbb R$. For a polynomial
external field of degree $2m$, the equilibrium measure is supported
on a union $S_V$ of at most $m$ disjoint intervals. Its density can
be written in the form \cite{DKM}
\begin{equation}\label{density}
\psi_V(s)=\prod_{j=1}^k \sqrt{(b_j-s)(s-a_j)}\ h(s),\qquad s\in
\cup_{j=1}^k[a_j,b_j],\ k\leq m,
\end{equation}
where $h$ is a polynomial of degree at most $2(m-k)$. The
equilibrium measure is characterized by the variational conditions
\begin{align}
&\label{var eq}2\int \log|s-y|d\mu(y)- V(s)=\ell_V,&
s\in\cup_{j=1}^k[a_j,b_j],\\
&\label{var ineq}2\int \log|s-y|d\mu(y)- V(s)\leq \ell_V,&
s\in\mathbb R.
\end{align}
The external field $V$ is called $k$-cut regular if $h(s)$ in
(\ref{density}) is strictly positive on $\cup_{j=1}^k[a_j,b_j]$ and
if (\ref{var ineq}) is strict for $s\in\mathbb R\setminus
\cup_{j=1}^k[a_j,b_j]$. In other words, it is singular if
\begin{itemize}
\item[(i)] equality in (\ref{var ineq}) holds at a point $s^*\in\mathbb R\setminus
\cup_{j=1}^k[a_j,b_j]$,
\item[(ii)] $h(s^*)=0$ with $s^*\in\cup_{j=1}^k(a_j,b_j)$.
\item[(iii)] $h(s^*)=0$ with $s^*=a_j$ or $s^*=b_j$.
\end{itemize}

\subsection{Example: quartic external field} Let us now study a
two-parameter family of quartic external fields
\begin{equation}
\label{V}
V_{x,t}(s)=e^x\left[(1-t)\frac{s^2}{2}+t\left(\frac{s^4}{20}-\frac{4s^3}{15}+\frac{s^2}{5}+\frac{8}{5}s\right)\right].
\end{equation}
For $t=0$, we have $V_{x,0}(s)=e^x\frac{s^2}{2}$, which means that
the random matrix ensemble is a rescaled Gaussian Unitary Ensemble.
The equilibrium measure $\mu_{x,0}$ is then given by
\begin{equation}
d\mu_{x,0}(s)=\frac{e^x}{2\pi}\sqrt{4e^{-x}-s^2}ds,\qquad
s\in[-2e^{-x/2},2e^{-x/2}].
\end{equation}
It can indeed be verified directly that this measure satisfies the
variational conditions (\ref{var eq})-(\ref{var ineq}). For $x=0$
and $0<t\leq 1$, one can verify that
\begin{equation}
d\mu_{0,t}(s)=\frac{1}{2\pi(5+\gamma^2)}\,\sqrt{s^2-4}((s-2)^2+\gamma^2)ds,\quad
s\in[-2,2],\quad \gamma=\sqrt{\frac{5}{t}-5}.
\end{equation}
This shows that $V_{0,1}$ has a singular point of type (iii) at
$s=2$. On the line $t=9$, $V_{x,9}(s)$ is symmetric around
$s^*=\frac{4}{3}$. The external field is one-cut regular for
$x<x^*=:-\log \frac{245}{9}$, and presumably two-cut for $x>x^*$. For $x\leq
x^*$, the equilibrium measure is given by
\begin{equation}
d\mu_{x,9}(s)=\frac{8}{\pi
b^2(b^2+4C)}\sqrt{(s-\frac{4}{3}+b)(\frac{4}{3}+b-s)}((s-\frac{4}{3})^2+C)ds,\qquad
s\in[s^*-b,s^*+b],
\end{equation}
where
\begin{equation}
\qquad b=\sqrt{\frac{140}{27} + \frac{4}{27} \sqrt{5} e^{-x}
\sqrt{27 e^x + 245 e^{2 x}}} \qquad C=\frac{e^{-x}}{36b^2}(80-9b^4e^x)
\end{equation}
At $x=x^*$, the equilibrium measure is given by
\begin{equation*}
d\mu_{x^*,9}(s)=\frac{8}{\pi
b^4}\sqrt{(s-\frac{4}{3}+b)(\frac{4}{3}+b-s)}(s-\frac{4}{3})^2ds,\qquad
s\in[\frac{4}{3}-b,\frac{4}{3}+b],\qquad b=\frac{2}{3}\sqrt{35},
\end{equation*}
which means that there is a type (ii) singular point at $s^*=4/3$.

\begin{figure}[t]
    \begin{center}
    \setlength{\unitlength}{1truemm}
    \begin{picture}(100,55)(-5,0)

    \put(50,0){\thicklines\circle*{.8}}
    \put(50,6){\thicklines\circle*{1.5}}
    \put(100,1){$x$}
    \put(52,67){$t$}
    \put(15,0){\vector(1,0){85}}
    \put(50,-3){\vector(0,1){65}}
    \put(100,55){$t=9$}
    \put(15,54){\line(1,0){85}}
    \put(30,54){\thicklines\circle*{.8}}
    \put(25,55){$(x^*,9)$}
     \qbezier(50,6)(40,7)(30,54)
    \qbezier(50,6)(80,52)(100,53)

    \put(20,20){1-cut}
    \put(80,20){1-cut}
    \put(52,36){2-cut}

    \put(28,37){(ii)}
    \put(69,37){(i)}
    \put(51,4){(iii)}
    \end{picture}
    \caption{Sketch of the phase diagram for the equilibrium measure in external field $V_{x,t}$.
    The one-cut region and the two-cut region are separated by two curves, at the left curve a
    type (ii) singular point is present, at the right curve a type (i) singular point,
    and at the intersection point $(0,1)$ there is a type (iii) singular point.}
    \label{Figure: RMTphasediagram}
\end{center}
\end{figure}
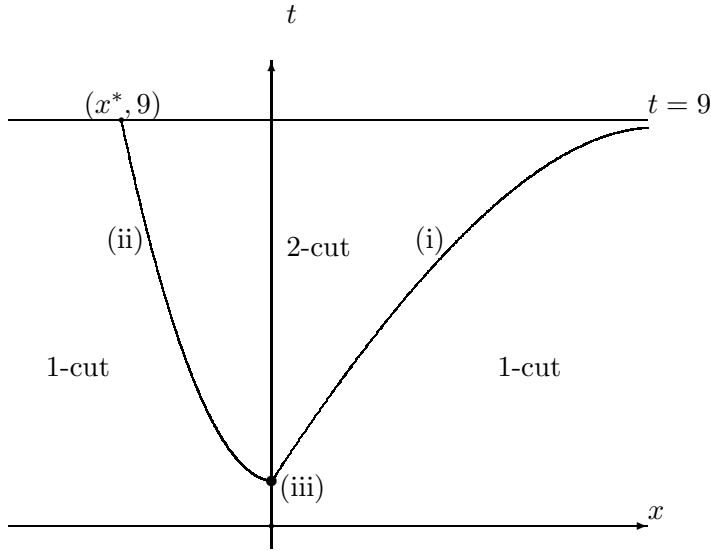

 For $t$
fixed and $x$ sufficiently large and positive, it follows from
results in \cite{KM} that the number of intervals is equal to the
number of global minima of $V_{x,t}$, which is one for $t<9$ and two
for $t=9$. For $t$ fixed and $x$ sufficiently large negative, one
can show that the equilibrium measure is supported on a single
interval. Also, for any $t$, when $x$ decreases, the support of the
equilibrium measure increases. This suggests that there are, as
shown in Figure \ref{Figure: RMTphasediagram}, two curves in the
$(x,t)$-plane where $V_{x,t}$ is singular: one connecting $(0,1)$
with $(x^*,9)$ where a singular point of type (ii) is present, and
one connecting $(0,1)$ with $(+\infty,9)$ where a singular point of
type (i) occurs.

\begin{remark}
In \cite{BertolaTovbis}, orthogonal polynomials with respect to complex weights of the form $e^{-nV(x)}$ were considered, with $V$ quartic symmetric with complex-valued leading coefficient. This lead to a phase diagram which shows certain similarities with ours, but also with breaking curves of a different nature.
\end{remark}

\subsection{Regular asymptotics}

If $V$ is a one-cut regular external field, the leading term of the
asymptotics for the recurrence coefficients depends in a very simple
way on the endpoints $a$ and $b$: we have \cite{DKMVZ2}
\begin{equation}\label{recursiean1}
         \gamma_{n} =
                \frac{b-a}{4}
                + \bigO(n^{-2}),\qquad \mbox{ as $n\to\infty$,}
    \end{equation}
    \begin{equation} \label{recursiebn1}
        \beta_{n} =
                \frac{b+a}{2}
                + \bigO(n^{-2}),\qquad\mbox{ as $n\to\infty$.}
    \end{equation}

If $V$ is a two-cut regular external field, the leading order term
in the asymptotic expansion for the recurrence coefficients is still
determined by the endpoints $a_1,b_1,a_2,b_2$, but the dependence is
somewhat more complicated, and the leading term is oscillating with
$n$. An explicit formula for the leading order asymptotics was given
and proved in \cite{DKMVZ2} for $k$-cut regular external fields $V$,
with $k$ arbitrary. We will not give details about those
asymptotics, but we note that the expansion is of a similar nature
as (\ref{elliptic1}) in the two-cut case.

\subsection{Critical asymptotics}

We will now describe the critical asymptotics for the recurrence
coefficients $\gamma_n(x,t)$ and $\beta_n(x,t)$ of the orthogonal
polynomials with respect to the weight $e^{-nV_{x,t}}$. It should be
noted that critical asymptotics near type (ii) and type (iii) singular points are known for more general
deformations of external fields $V_{x,t}$ than only the one defined
by (\ref{V}).

\subsubsection{Singular interior points}
Assume that $V_{x^*,t^*}(s)$ is a singular external field with a
singular point $s^*$ of type (ii) (a singular interior point), and
with support $[a,b]$ of the equilibrium measure. Asymptotics for the
recurrence coefficients were obtained in \cite{BI} for quartic
symmetric $V$ and in \cite{CKV} for real analytic $V$. Let us
specialize the results to our example where $V_{x,t}$ is given by
(\ref{V}). Since $V_{x^*,t^*}$ is quartic, this implies that
$\psi_{x^*,t^*}$ has the form
\begin{equation}
\psi_{x^*,t^*}(s)=C\sqrt{(s-a)(b-s)}(s-s^*)^2.
\end{equation}
Then as $n\to\infty$ simultaneously with $x \to x^*$ such that
    $x-x^* = \bigO(n^{-2/3})$, we have the asymptotic expansions
    \cite{CKV}
    \begin{align} \label{recuran}
        &\gamma_{n}(x,t^*) =
                \frac{b-a}{4}
                - \frac{1}{2c}q\left(s_{x,n}\right)\cos(2\pi n\omega(x))
                n^{-1/3}
                + \bigO(n^{-2/3}),\\&\label{recurbn}
        \beta_{n}(x,t^*) =
                \frac{b+a}{2}
                + \frac{1}{c}q\left(s_{x,n}\right)\sin(2 \pi n\omega(x)+
                \theta) n^{-1/3}
                + \bigO(n^{-2/3}),
    \end{align}
    where
    \[s_{x,n}=n^{2/3}(e^{x^*-x}-1)\frac{1}{c \sqrt{(s^*-a)(b-s^*)}},\]
    and
    where $c$, $\theta$, and $\omega$ are given by
    \begin{align*}&c=\left(\frac{\pi C\sqrt{(s^*-a)(b-s^*)}}{4}\right)^{1/3}, \qquad
    \theta = \arcsin \frac{b+a}{b-a},\\&
        \omega(x) = \int_0^{b}\psi_{x^*,t^*}(s)ds+\bigO(n^{-2/3}),\qquad \mbox{ as $n\to\infty$}.
    \end{align*} An exact formula for $\omega$ can be given in
    terms of a modified equilibrium problem.
 When $x$ approaches $x^*$, we observe that the
recurrence coefficients develop oscillations. The envelope of the
oscillations is described by the Hastings-McLeod solution $q$. One
should compare formulas (\ref{recuran})-(\ref{recurbn}) with
(\ref{expansionu}) and note that the scalings correspond after
identifying $\epsilon$ with $1/n$.

\subsubsection{Singular edge points}

Asymptotics for the recurrence coefficients for general one-cut
external fields $V$ with a singular endpoint were obtained in
\cite{CV2}. Let $V_0$ be an external field such that the equilibrium
measure is supported on $[a,b]$ and such that the density $\psi_0$
behaves like $\psi_0(s)\sim c(b-s)^{5/2}$ as $s\to b$, $c\neq 0$.
Double scaling asymptotics were obtained for external fields of the
form $V_0+SV_1+TV_2$  with   real $S,T\to 0$, where $V_1$ is arbitrary and
$V_2$ satisfies the condition
\[\int_a^b \sqrt{\frac{s-a}{b-s}}V_2'(s)ds=0.\]
We can write $V_{x,t}$ in the form
\begin{equation}
V_{x,t}(s)=V_{0,1}(s)+(e^x-1)V_{0,1}(s)+e^x(t-1)(V_{0,1}(s)-V_{0,0}(s)).
\end{equation}
Since
\[\int_{-2}^2 \sqrt{\frac{s+2}{2-s}}(V_{0,1}'(s)-V_{0,0}'(s))ds=0,\]
we can apply the results of \cite{CV2}. In the double
    scaling limit where $n\to\infty$ and simultaneously $x\to 0$, $t\to 1$ in such a way that
    $\lim n^{6/7}(e^x-1)$ and $\lim n^{4/7}e^x(t-1)$ exist,
    we have
    \begin{align}\label{recurrenceedge1}
        &\gamma_n(x,t) =1+\frac{1}{2c}\,
        U(c_1n^{6/7}(e^x-1),c_2n^{4/7}e^x(t-1))n^{-2/7}+\bigO(n^{-4/7}),\\&\label{recurrenceedge2}
        \beta_n(x,t) =\frac{1}{c}\,
        U(c_1n^{6/7}(e^x-1),c_2n^{4/7}e^x(t-1))n^{-2/7}+\bigO(n^{-4/7}).
    \end{align}
    The constants $c, c_1, c_2$ are given by
    \begin{align*}\label{definition: cc1c2}
    &c=6^{2/7}> 0,\\
    &c_1=\frac{1}{2\pi c^{1/2}}\int_{-2}^2
    \sqrt{\frac{u-2}{2-u}}V_{0,1}'(u)du=6^{-1/7},\\
 &c_2=\frac{1}{4\pi i c^{3/2}}\int_\gamma
\sqrt{\frac{2+u}{(2-u)^3}}(V_{0,1}'(u)-V_{0,0}'(u))du=2.6^{-3/7},
\end{align*}
where $\gamma$ is a counterclockwise oriented contour encircling
$[-2,2]$.
\begin{remark}Applying the results from \cite{CV2} directly, one has
an error term $\bigO(n^{-3/7})$ in (\ref{recurrenceedge1}) and
(\ref{recurrenceedge2}), but going through the calculations, it can
be verified that the error term is actually $\bigO(n^{-4/7})$. The
analogy between (\ref{recurrenceedge1})-(\ref{recurrenceedge2}) and
(\ref{expansionu}) is obvious.
\end{remark}

\subsubsection{Singular exterior points}

Asymptotics for the recurrence coefficients in the vicinity of a
singular exterior point have not appeared in the literature to the
best of our knowledge. Asymptotics for orthogonal polynomials
associated to an external field $V$ with a singular exterior point
and for the correlation kernel (\ref{kernel}) have been studied in
\cite{C,BL,M} using the Riemann-Hilbert approach. We are convinced
that the same analysis can be used, with some additional effort, to
compute asymptotics for the recurrence coefficients. If
$V_{x^*,t^*}$ is an external field with a singular exterior point,
the analogy with the KdV asymptotics suggests asymptotic expansions
of the form
\begin{align}
&        \gamma_n(x^* - y\frac{\ln n}{c_0n} ,t)
=\frac{b(x^*,t)-a(x^*,t)}{4}+
        c_1 \sum_{k=0}^{\infty}\sech^2(X_k)+\bigO(n^{-1}\ln^2 n)
 \\&     \beta_n(x^* - y\frac{\ln n}{c_0n} ,t) =\frac{b(x^*,t)+a(x^*,t)}{2}+
        c_1 \sum_{k=0}^{\infty}\sech^2(X_k)+\bigO(n^{-1}\ln^2 n),
    \end{align}
    as $n\to\infty$,
where
\[
X_k=-c_2(y,k)\ln n + c_3(k).\]

\section{The problem of matching}

Asymptotic expansions for KdV solutions are known in the regular
regions and in critical regions, but we do not have uniform
asymptotics for $u(x,t,\e)$ in $x$ and $t$. Indeed, the critical
asymptotics are only valid in shrinking neighborhoods of the
breaking curves: a neighborhood of size $\bigO(\e^{2/3})$ near the
leading edge, a neighborhood of size $\bigO(\e\ln\e)$ near the
trailing edge, and a neighborhood of size $\bigO(\e^{4/7})$ at the
point of gradient catastrophe. On the other hand, the regular
asymptotics are only proved to hold uniformly for $x$ and $t$ at a
fixed distance away from the breaking curves. However one can see
easily that (\ref{expansionu}) and (\ref{expansion u}) match
formally with the regular asymptotics for $x$ close to the breaking
curves but outside the cusp-shaped region. Indeed for
(\ref{expansionu}) this follows from the decay of the
Hastings-McLeod solution $q$ at$+\infty$. When $x$ is close to the
boundary but inside the cusp-shaped region, the situation is more
complicated. One can hope that the regular asymptotics can be
improved in such a way that they hold also when $x,t$ approach a
breaking curve sufficiently slowly when $\epsilon$ tends to $0$, and
that the critical asymptotics can be improved to hold in a slightly
bigger neighborhood of the breaking curves. It would be of interest
to see if such an approach could provide uniform asymptotics for the
KdV solution as $\epsilon\to 0$.

\medskip

The problem of obtaining uniform asymptotics in $x$ and $t$ for the
recurrence coefficients $\gamma_n(x,t)$ and $\beta_n(x,t)$ may seem
an artificial one at first sight, since one is often interested in a
random matrix with a fixed external field $V$ instead of letting $V$
vary. However, it becomes more relevant when studying the partition
function
\[Z_n=\int_{\mathbb R^n}\prod_{i<j}(\lambda_i-\lambda_j)^2\prod_{i=1}^ne^{-nV(\lambda_i)}d\lambda_i.\]
It is well-known that \[Z_n=n!\prod_{j=1}^{n-1} \kappa_j^{-2},\]
where $\kappa_j$ is the leading coefficient of the normalized
orthogonal polynomial $p_j$ with respect to the weight $e^{-nV}$. A
consequence of this formula is that, if one lets $V$ vary with a
parameter $\tau$ in a convenient way, it is possible to derive
various identities for $\tau$-derivatives of $\ln Z_n$ in terms of
the recurrence coefficients $\gamma_k(\tau)$ and $\beta_k(\tau)$ for
$k$ large \cite{BI3, EMcL}. A possible strategy to obtain
asymptotics for the partition function, is to let the
$\tau$-dependence be such that $V$ interpolates between the Gaussian
$V(z;\tau_0)=\frac{z^2}{2}$ and $V(z;\tau_1)=V(z)$. Integrating the
differential identity then requires asymptotics for the Gaussian
partition function (which are known) and uniform asymptotics for the
recurrence coefficients $\gamma_n(\tau)$ and $\beta_n(\tau)$ over
the whole range $[\tau_0,\tau_1]$. Depending on the chosen deformation,
this could require uniform asymptotics for the recurrence
coefficients near a singular point of type (i), (ii), or (iii). The
results presented in the previous section do not provide
sufficiently detailed asymptotics for the recurrence coefficients:
they are not uniform near the breaking curves. For example near a
critical point of type (ii), formulas
(\ref{recuran})-(\ref{recurbn}) are only valid for
$x-x^*=\bigO(n^{-2/3})$ as $n\to\infty$, whereas the asymptotic
formula in the two-cut region is valid only at a fixed distance away
from a critical point.

\section{The Toda lattice and KdV}

It is well-known that recurrence coefficients for orthogonal
polynomials follow the time flows of the Toda hierarchy. In this
section, following \cite{Dubrovin3, DGKM} we will formally derive the
KdV equation as a scaling limit of the continuum limit of the Toda
lattice. This gives a heuristic argument why asymptotics for KdV and
the recurrence coefficients show similarities.

The Toda lattice is a Hamiltonian system described by the equations
\begin{equation}
\label{Toda} \dfrac{d u_n}{dt}=v_n-v_{n-1},\quad \dfrac{d
v_n}{dt}=e^{u_{n+1}}-e^{u_n},\quad n\in\mathbb{Z}.
\end{equation}
The Toda lattice is a prototypical example of a completely
integrable system \cite{Flaschka}.
Let
\begin{equation}\label{f2}
V(\xi) = V_0(\xi)+\sum_{j=1}^{2d}t_j\xi^{j}, \quad t_{2d} >0,
\end{equation}
where $V_0(\xi)$ is a fixed polynomial of even degree with positive
leading coefficient, and let $p_j$ be the orthogonal polynomials
defined by
\begin{equation}\label{f5}
\int_{-\infty}^{\infty}p_{n}(\xi)p_{m}(\xi)e^{-\frac{1}{\epsilon}V(\xi)}\,d\xi
=\delta_{nm}\,,
\end{equation}
where $\epsilon=\dfrac{1}{N}$ is a small positive parameter. As mentioned before,
the polynomials $p_n(\xi)$ satisfy a three term recurrence relation
of the form (\ref{recur}).

The recurrence coefficients $\gamma_n$ and $\beta_n$  in (\ref{recur}) evolve with
respect to the times $t_k$ defined in (\ref{f2})  according to the equations \cite{Eyn, DS,
FIK, BEH}
\begin{align}
\label{prop12} \epsilon\frac{\partial\gamma_{n}}{\partial t_k}
&= \frac{\ga_n}{2}\left([Q^k]_{n-1,n-1}-[Q^k]_{nn}\right),\\
\label{prop13} \epsilon\frac{\partial\beta_{n}}{\partial t_k} &=
\gamma_n[Q^k]_{n,n-1}-\gamma_{n+1}[Q^k]_{n+1,n},
\end{align}
where $[Q^k]_{n,m}$ denotes the $n,m$-th element of the matrix $Q^k$
and $Q$ is the tridiagonal matrix
\begin{equation}
Q=
\begin{pmatrix}
\be_0 & \ga _1 & 0 & 0 & 0 & \dots \\
\ga_1 & \be_1 & \ga _2 & 0 & 0 & \dots  \\
0 & \ga_2 & \be_2 & \ga _3 & 0 & \dots  \\
0 & 0 & \ga_3 & \be_3 & \ga_4 & \dots  \\
0 & 0 & 0 & \ga_4 & \be_4 & \dots  \\
\vdots & \vdots & \vdots & \vdots & \vdots & \ddots
\end{pmatrix}.
\end{equation}
The equations (\ref{prop12})-(\ref{prop13}) are the Toda lattice
hierarchy in the Flaschka variables \cite{Flaschka}. In particular the first flow of
the hierarchy takes the form
\begin{equation}\label{Todafirst}
\begin{split}
&\epsilon\frac{\partial\gamma_n}{\partial t_1} =
\frac{\ga_n}{2}\left(\beta_{n-1}-\beta_{n}\right),\\
&\epsilon\frac{\partial\be_n}{\partial t_1} = \ga_n^2-\ga^2_{n+1}.
\end{split}
\end{equation}
These equations correspond to the Toda lattice (\ref{Toda}) by
identifying $t_1=t$ , $\beta_n=-v_n$ and
\begin{equation}
u_n=\log\gamma_n^2.
\end{equation}
In addition to the Toda equations, the recurrence coefficients for
the orthogonal polynomials satisfy a constraint that is given by the
discrete string equation which takes the form  \cite{FIK}
 \begin{equation}
 \begin{split}
 \label{string}
 &\gamma_n[V'(Q)]_{n,n-1}=n\epsilon,\\
 &[V'(Q)]_{n,n}=0.
 \end{split}
 \end{equation}
 For example, choosing $V_0(\xi)=\frac{1}{2}\xi^2$ one obtains
 \begin{equation}
 \label{initialdata}
 \beta_n(\boldsymbol{t}=0)=0,\quad \gamma_n^2(\boldsymbol{t}=0)=n\epsilon,\quad \boldsymbol{t}=(t_1,t_2,\dots,t_{2d}).
 \end{equation}

To obtain the  continuum limit of the Toda lattice, let us assume
that $u(x)$ and $v(x)$ are smooth functions that interpolate the
sequences $u_n,v_n$ in the following way: $u(\epsilon n)=u_n$ and
$v(\epsilon n)=v_n$ for some small $\epsilon>0$, $n>0$, $x=\epsilon n$. Then the
Toda lattice (\ref{Toda}) reduces to an evolutionary PDE of the form
\cite{EY, DeiftMcLaughlin}
\begin{equation}
\label{Todac}
\begin{split}
&u_t=\dfrac{1}{\epsilon}\left[v(x)-v(x-\epsilon)\right]=v_x-\dfrac{1}{2}\epsilon v_{xx}+O(\epsilon^2) \\
&v_t=\dfrac{1}{\epsilon}\left[e^{u(x+\epsilon)}-e^{u(x)}\right]=e^uu_x+\dfrac{1}{2}\epsilon
(e^u)_{xx}+O(\epsilon^2).
\end{split}
\end{equation}

In order to write the continuum limit of the Toda lattice in a
canonical Hamiltonian form,  following  Dubrovin-Zhang \cite{DZ},
we introduce $w(x)$  by
 \begin{equation}
 \label{canonical}
 w(x)=\epsilon\partial_x[1-e^{-\epsilon\partial_x}]^{-1}u(x)=w+\dfrac{\epsilon}{2} w_x+\dfrac{\epsilon^2}{12}w_{xx}+\dots, \end{equation}

 In the coordinates $v,w$ the  continuum limit of the Toda lattice equations takes the form
 \begin{equation}
 \label{Todacc}
 \begin{split}
& w_t=v_x\\
&v_t=e^w\left[w_x+\dfrac{\epsilon^2}{24}(2w_{xxx}+4w_xw_{xx}+w_x^3)\right]+O(\epsilon^4)
 \end{split}\end{equation}
 with the corresponding Hamiltonian given by
 $H=\int\left[\dfrac{v^2}{2}+e^w-\dfrac{\epsilon^2}{24}e^ww_{x}^2+\dots\right]dx$ and Poisson bracket $\{v(x),w(y)\}=\delta'(x-y)$ where $\delta(x)$ is the Dirac $\delta$ function.
 We remark that in these coordinates the continuum limit  of the Toda equation contains only even terms in $\epsilon$.
 For $\epsilon=0$, (\ref{Todacc}) reduces to
 \begin{equation}
 \label{Todad}
 w_t=v_x,\quad v_t=e^ww_x.
 \end{equation}
 The solution of  equations (\ref{Todad}) can be obtained by the method of characteristics.
 The initial data relevant to us should satisfy the  continuum limit of the string equation (\ref{string}) for
 $t=0$.
 The Riemann invariants of (\ref{Todad})  are
 \[
 r_{\pm}=v\pm 2e^{\frac{w}{2}}
 \]
 so that (\ref{Todad}) takes the form
 \[
 \dfrac{\partial}{\partial t}r_{\pm}+\lambda_{\pm}\dfrac{\partial}{\partial x}r_{\pm}=0,\quad \lambda_{\pm}=\mp e^{\frac{w}{2}}=\mp\dfrac{r_+-r_-}{4}.
 \]
 The generic solution of (\ref{Todad})  can be written in the form \cite{Tsarev, W}
 \begin{equation}
 \label{hodograph}
 x=\lambda_{\pm} t+f_{\pm}(r_+,r_-),
 \end{equation}
 where  $f_{\pm}(r_+,r_-)$ are two  functions
that satisfy the equations \cite{Tsarev}
\begin{equation}
\label{ED} \dfrac{\partial}{\partial r_-}f_{+}= \dfrac{\partial
\lambda_+}{\partial
r_-}\dfrac{f_+-f_-}{\lambda_+-\lambda_-}=-\dfrac{f_+-f_-}{2(r_+-r_-)}=\dfrac{\partial}{\partial
r_{+}}f_{-}.
\end{equation}
From the above relations one can conclude that there exists a
function $f=f(r_+,r_-)$ so that
\[
f_{\pm}=\dfrac{\partial f}{\partial r_{\pm}}.
\]
The explicit  dependence of $f$ for a certain class of  initial data
can be found in \cite{DeiftMcLaughlin}. To obtain $f$ in the random
matrix case we impose that the  equations (\ref{hodograph}) are
consistent with the continuum limit of the discrete  string equation
(\ref{string}) for $t_1=t\geq 0$ and $t_j=0$ for $j>1$. At the
leading order in $\epsilon$ the string equation (\ref{string}) in
the Riemann invariants $r_{\pm}=-\beta\pm 2\gamma$ gives after
straightforward but long calculations,  the following  expression
for the function $f(r_+, r_-)$:
\begin{equation}
\label{f}
f(r_+,r_-)=-\mbox{Res}_{\xi=\infty}\left[V_0'(\xi)\sqrt{(\xi-r_+)(\xi-r_-)}d\xi\right].
\end{equation}
\begin{remark}
The equations (\ref{hodograph}) with $f$ given in (\ref{f}),
coincide with the equations that define the support of the
equilibrium measure for the variational problem
\[
\inf\limits_{\int_{\mathbb{R}}
d\nu(\xi)=1}\left[\int_{\mathbb{R}}\int_{\mathbb{R}}\log\frac{1}{|\xi-\eta|}
d\nu(\xi)d\nu(\eta)+\dfrac{1}{x}\int_{\mathbb{R}}
V(\xi)d\nu(\xi)\right]
\]
in the case where the equilibrium measure is supported on one
interval. The Riemann invariants $r_+$ and $r_-$ can thus be
interpreted as the end-points of the support of the equilibrium
measure.
\end{remark}
In what follows, we are going to show that the  solution  of the
equation (\ref{Todacc}) in  the vicinity  of a singular point of
type (iii)  reduces to the KdV equation, in agreement with
\cite{Dubrovin3}. First  we consider the solution of the hodograph
equation (\ref{hodograph}) near a singular point of type (iii);
namely, let $(x_c,t_c)$ be a point of gradient catastrophe for the
Riemann invariant $r_+$, which means that $\partial_xr_+$ goes to
infinity at the critical point $(x_c,t_c)$. We define
$r_{\pm}(x_c,t_c)=r_{\pm}^c$. Such a critical point is characterized
by the conditions
\[
\lambda^c_{+,+}t_c+f^c_{+,+}=0,\quad
\lambda^c_{+,++}t_c+f^c_{+,++}=0,
\]
and the critical  point is generic if
\[
\lambda^c_{+,+++}t_c+f^c_{+,+++}\neq 0,\quad
\lambda^c_{-,-}t_c+f^c_{-,-}\neq 0,
\]
where in the above formulas we used the notation
$\lambda^c_{-,-}=\dfrac{\partial}{\partial
r_-}\lambda_-(r_+=r_+^c,r_-=r_-^c)$ and consistently for the other
terms.

Expanding in power series (\ref{hodograph})  near $(x_c,t_c)$  and
using (\ref{ED}) after the rescalings
\begin{equation}
\label{scalings}
\begin{split}
&x_-=k^{-2/3}(x-x_c-\lambda_-^c(t-t_c)),\quad x_+=k^{-1}(x-x_c-\lambda_+^c(t-t_c))\\
&\bar{r}_-=k^{-2/3}(r_--r_-^c),\quad \bar{r}_+=k^{-1/3}(r_+-r_+^c),
\end{split}
\end{equation}
one obtains, letting $k\rightarrow 0$,
\begin{equation}
\label{Withney}
\begin{split}
&x_-=c_1\bar{r}_-\\
&x_+=c_2 x_-\bar{r}_++c_3\bar{r}^3_+,
\end{split}
\end{equation}
where
\begin{equation}
\label{cs} c_1=(f_{-,-}^c+\lambda_{-,-}^ct_c),\quad
c_2=\dfrac{\lambda_{+,+}^c}{\lambda_+^c-\lambda_-^c},\;\;
c_3=\dfrac{1}{6}(\lambda_{+,+++}^ct_c+f_{+,+++}^c).
\end{equation}
We observe that (\ref{Withney}) describes a Withney singularity in
the neighbourhhood of  $(0,0)$ \cite{Dubrovin3}. Performing the same
rescalings (\ref{scalings})  to the equations (\ref{Todad}) and
letting $k\rightarrow 0$ one obtains
\[
\dfrac{\partial \bar{r}_-}{\partial x_+}=0,\;\;\;\dfrac{\partial
\bar{r}_+}{\partial x_-}+c_2 \bar{r}_+\dfrac{\partial
\bar{r}_+}{\partial x_+}=0,
\]
with $c_2$ as in (\ref{cs}).  Clearly the equations (\ref{Withney})
are a solution  of the above equations  with  singularity  in $(x_+=0,x_-=0)$ and at
$\bar{r}_{\pm}=0$. The next step is to
perform the rescaling (\ref{scalings}) to the  equation
(\ref{Todacc}) and letting $\epsilon\rightarrow
k^{\frac{7}{6}}\epsilon$. One obtains in the limit $k\rightarrow 0$
\begin{equation}
\begin{split}
&\bar{r}_-=\dfrac{x_-}{c_1}+c_4\epsilon^2\dfrac{\partial^2}{\partial x_+^2}\bar{r}_+,\;\;\;c_4=\dfrac{r_+^c-r_-^c}{192(\lambda_-^c-\lambda_+^c)}=\dfrac{1}{96}\\
&\dfrac{\partial \bar{r}_+}{\partial x_-}+c_2
\bar{r}_+\dfrac{\partial\bar{ r}_+}{\partial
x_+}+c_4\epsilon^2\dfrac{\partial^3}{\partial x_+^3}\bar{r}_+=0.
\end{split}
\end{equation}
The first of the above equations has been obtained after integration
with respect to $x_+$ using (\ref{Withney}). The  second one is the
KdV equation for $r_+$ with time  variable  $x_-$ and space variable
$x_+$. Such derivation has been obtained in a more general setting
in \cite{DGKM}. On the formal level, the above calculations explain
why the asymptotic behavior of  the solution of the continuum limit
of Toda lattice and in particular of the recurrence coefficients of
orthogonal polynomials  near the point of gradient catastrophe  is
of a similar nature as the KdV case. However, a rigorous proof of
the generic behavior of the solution of the continuum limit of Toda
lattice near the point of gradient catastrophe cannot be derived
from the KdV case but a separate proof is needed.

\section*{Acknowledgements}
The authors acknowledge support by ERC Advanced Grant FroMPDE. TC was also supported by FNRS, by the Belgian Interuniversity Attraction Pole P06/02, P07/18 and by the European Research Council under the European Union's Seventh Framework Programme (FP/2007/2013)/ ERC Grant Agreement n. 307074.

\end{document}